\documentclass[3p]{elsarticle}
\UseRawInputEncoding
\usepackage{lineno,hyperref}
\usepackage{graphicx}
\usepackage{amsmath}
\usepackage{color}
\usepackage[normalem]{ulem}
\usepackage{mathtools}

\usepackage{rotating}
\usepackage{lscape}

\journal{arXiv}
\begin{document}

\begin{frontmatter}

\title{Rethinking Generalized Beta Family of Distributions}

\author[mymainaddress]{Jiong Liu}
\author[mymainaddress]{R. A. Serota\fnref{myfootnote}}
\fntext[myfootnote]{serota@ucmail.uc.edu}

\address[mymainaddress]{Department of Physics, University of Cincinnati, Cincinnati, Ohio 45221-0011}

\begin{abstract}
We approach the Generalized Beta (GB) family of distributions using a mean-reverting stochastic differential equation (SDE) for a power of the variable, whose steady-state (stationary) probability density function (PDF) is a modified GB (mGB) distribution. The SDE approach allows for a lucid explanation of Generalized Beta Prime (GB2) and Generalized Beta (GB1) limits of GB distribution and, further down, of Generalized Inverse Gamma (GIGa) and Generalized Gamma (GGa) limits, as well as describe the transition between the latter two. We provide an alternative form to the "traditional" GB PDF to underscore that a great deal of usefulness of GB distribution lies in its allowing a long-range power-law behavior to be ultimately terminated at a finite value. We derive the cumulative distribution function (CDF) of the "traditional" GB, which belongs to the family generated by the regularized beta function and is crucial for analysis of the tails of the distribution. We analyze fifty years of historical data on realized market volatility, specifically for S\&P500, as a case study of the use of GB/mGB distributions and show that its behavior is consistent with that of negative Dragon Kings.
\end{abstract}


\end{frontmatter}

\section{Introduction\label{Intro}}

Generalized Beta family of distributions has been widely used in econometrics and econophysics, in particular for understanding of distributions of income and wealth \cite{mcdonald1984some, mcdonald1995generalization, chotikapanich2008modelling, chakrabarti2013econophysics, biewen2018econometrics}\footnote{There exist numerous  extensions  of GB as well -- see e.g. \cite{gomez2018family}}. Family members with scale-free, power-law tails may be of particular interest in this regard \cite{chotikapanich2018using}, which prompted development of stochastic models of economic exchange that produced power-law-tailed IGa \cite{bouchaud2000wealth}, GIGa \cite{ma2013distribution}, and GB2 \cite{dashti2020stochastic} as their steady-state distributions. These distributions have other numerous applications in various fields, such as physics \cite{sepplinen2012scaling, thiery2014loggamma, grange2017loggamma}, mathematical finance \cite{praetz1972distribution, cox1985theory, nelson1990arch, heston1993closed, dragulescu2002probability, fuentes2009universal, ma2014model}, and cognitive psychology \cite{dashti2020modeling}, to name a few. Of particular importance to this work is that a very general yet simple model of stochastic volatility yields a GB2 steady-state distribution \cite{dashti2021combined}.

The first main objective of this work was to generalize GB distribution itself, relative to the one introduced in \cite{mcdonald1995generalization}, by removing the correlations between the two scale parameters. The main purpose of such generalization is to obtain a distribution which has long power-law tails above the first scale, only to be eventually terminated, via a drop-off to zero, at the second scale. The need for such distribution was motivated by our looking into a possibility of Dragon Kings (DK) \cite{sornette2009,sornette2012dragon} in the realized volatility (RV) \cite{dashti2021realized}. In particular, it is of great interest to glean into whether the most calamitous events in the stock market -- Savings and Loan Crisis, Tech Bubble, Great Recession and Covid Pandemic -- follow the Black Swan (BS) behavior and stay on the power-law tails or they deviate strongly higher and thus are DK. What we found instead \cite{liu2022dragon} was behavior consistent with \emph{negative} Dragon Kings (nDK) \cite{pisarenko2012robust}, which are a strong drop-offs of the power-law tails, and are well described by GB-type distribution functions, mGB.

While the most general description of stochastic dynamic models for Generalized Beta family of distributions can be found in \cite{hertzler2003classical}, the model for the top GB -- or for that matter the top Beta (B) distribution -- is notably absent. Therefore, deriving such model was the second main objective of this work. As it turns out, the "traditional" GB distribution -- as defined in \cite{mcdonald1995generalization} and generalized here -- does not appear to be a steady-state solution of an SDE. However, we did identify an SDE whose steady-state distribution -- modified Beta (mB) -- is very similar in properties to the Beta (B) distribution. We then constructed an mGB from mB using a change of variable to the power of the variable. We also derived an SDE from the mB SDE via the same change of variable, which yielded yet another mGB as its steady-state distribution. Both mGB distributions have key features of the traditional GB, and its generalization here, but seem to describe RV distribution better. There is very little numerical difference between the two mGB distributions, except that the former has a far simpler analytical form than the latter.

This paper is organized as follows. In Sec. \ref{GB} we discuss the generalization of the GB distribution that is cast in terms of two scale and three shape parameters, with one of the shape parameters representing a change of variable to the power of the variable, the power being the said shape parameter; this change of variable also signifies transition from B to GB distribution. We also derive CDF of GB distribution and show that it belongs to a class of incomplete-Beta-function-generated distributions \cite{eugene2002betanormal, jones2004families, cordeiro2009new, alexander2012generalized, alzaatrech2013newmethod, lemonte2013extended}. In Sec.\ref{SDE} we derive two novel SDE that produce mB and two mGB distributions and explore their properties relative to the "traditional" GB distribution. We also develop an SDE framework of producing GB family hierarchy and explain the nature of transition between power-law-tailed distributions and those with exponential-like tails. In Sec. \ref{Volatility} we investigate distributions of S\&P realized volatility vis-a-vis fits by GB and mGB distributions and show that mGB provides a better fit and that termination of the distribution at a finite values is consistent with nDK behavior. We summarize our results in Sec. \ref{Summary}.

\section{Generalized Beta Distribution\label{GB}}

Employing a "dimensionless" variable, we use the following form of the GB PDF:

\begin{equation}
f_{GB}(x;\alpha,\beta _1,\beta _2, p,q)=\frac{\alpha}{\beta _1 B(p,q)} \left(1+\left(\frac{\beta _1}{\beta _2}\right)^{\alpha }\right)^p \left(\frac{x}{\beta _1}\right)^{\alpha  p-1} \left(1-\left(\frac{x}{\beta _1}\right)^{\alpha }\right)^{q-1} \left(1+\left(\frac{x}{\beta _2}\right)^{\alpha}\right)^{-p-q},
\label{GBPDF}
\end{equation}
where $\beta _1$ and $\beta _2$ are scale parameters and $\alpha$, $p$ and $q$ are shape parameters, all positive, $B(p,q)$ is the beta function and $x \leq \beta _1$. This transforms into the form used by McDonald \cite{mcdonald1995generalization} with the substitution $\beta _1=\beta/(1-c)^{1/\alpha}$ and $\beta _2=\beta/c^{1/\alpha}$, $0\leq c \leq1$, however we do not find it necessary to impose such correlations between $\beta _1$ and $\beta _2$. We derive the following expressions for GB CDF and 1-CDF, the latter being important in investigating power-law dependencies:

\begin{equation}
F_{GB}(x;\alpha,\beta _1,\beta _2, p,q)=I\left({\frac{\left(\frac{x}{\beta _1}\right)^{\alpha }+\left(\frac{x}{\beta _2}\right)^{\alpha }}{1+\left(\frac{x}{\beta _2}\right)^{\alpha }}};p,q\right),
\label{GBCDF}
\end{equation}

\begin{equation}
1-F_{GB}(x;\alpha,\beta _1,\beta _2, p,q)=I\left({\frac{1-\left(\frac{x}{\beta _1}\right)^{\alpha }}{1+\left(\frac{x}{\beta _2}\right)^{\alpha }}};q,p\right),
\label{GBCDFtail}
\end{equation}
where $I(y;p,q)=B(y;p,q)/B(p,q)$ and $B(y;p,q)$ are, respectively, the regularized and incomplete beta functions \cite{nist2022digital}. 

Eqs. (\ref{GBPDF})-(\ref{GBCDFtail}) reproduce well-known GB hierarchies \cite{mcdonald1984some, mcdonald1995generalization, hertzler2003classical, chotikapanich2008modelling}: GB1 and GB2 for $\beta_2\to +\infty$ and $\beta_1\to +\infty$ respectively, and further B1 and B2 (Beta Prime) for $\alpha=1$. We arrive at the latter two also by taking the limits in reverse order: first $\alpha=1$, which yields Beta (B) distribution and then $\beta_2\to +\infty$ or $\beta_1\to +\infty$. The further down limits of GGa and GIGa are best understood from analysis of an SDE that produces all the above distributions and will be discussed in Sec. \ref{SDE}. We also observe the special nature of the shape parameter $\alpha$ in that GB, and all its family members, are easily generated form the B family ($\alpha=1$) by a simple change of variable 
\begin{equation}
f_{B}(y; \beta _1,\beta _2, p,q) \, dy = f_{GB}(x;\alpha,\beta _1,\beta _2, p,q) \, dx
\label{B-GB}
\end{equation}
 where $y=x^{\alpha}$ and
 \begin{equation}
f_{B}(x;\beta _1,\beta _2, p,q)=\frac{1}{\beta _1 B(p,q)} \left(1+\frac{\beta _1}{\beta _2}\right)^p \left(\frac{x}{\beta _1}\right)^{p-1} \left(1-\frac{x}{\beta _1}\right)^{q-1} \left(1+\frac{x}{\beta _2}\right)^{-p-q},
\label{BPDF}
\end{equation}
with the appropriate rescaling of $\beta_1$ and $\beta_2$. The same holds true also for the corresponding SDE in Sec. \ref{SDE}, where the mean reversion SDE for variable $x$ produces B family of steady-state distributions, while the same mean-reversion SDE for variable $x^\alpha$ produces GB family. More precisely, the steady-state distribution of the corresponding SDE for the variable $x$ is a modified Beta distribution, mB, which is very similar to B, while the steady--state distribution for variable  $x^\alpha$ is a modified GB, mGB. One can also obtain a second modified GB distribution via the change of variable (\ref{B-GB}) in mB. Again, both mGB are very similar to GB. It should be also emphasized that below B/GB level al members of the respective families can be written in a traditional form - see footnote \ref{SDEcorrespondence} in Sec. \ref{SDE} below.

It should be noted that since the variable of the regularized beta changes between $0$ and $1$, GB CDF belongs to a class of distributions generated by a seed CDF \cite{eugene2002betanormal, jones2004families, cordeiro2009new, alexander2012generalized, alzaatrech2013newmethod, lemonte2013extended}, see also \cite{srinivasa2010distance}, which in this case is given by

\begin{equation}
F(x;\alpha,\beta _1,\beta _2)={\frac{\left(\frac{x}{\beta _1}\right)^{\alpha }+\left(\frac{x}{\beta _2}\right)^{\alpha }}{1+\left(\frac{x}{\beta _2}\right)^{\alpha }}}.
\label{Fseed}
\end{equation}
In the limiting cases of GB1, $\beta_2 \rightarrow \infty$, and GB2, $\beta_1 \rightarrow \infty$, it reproduces the results of \cite{sepanski2007family}. The corresponding PDF $f(x;\alpha,\beta _1,\beta _2)=F'(x;\alpha,\beta _1,\beta _2)$, GB PDF can be rewritten as \cite{eugene2002betanormal}

\begin{equation}
f_{GB}(x;\alpha,\beta _1,\beta _2, p,q)=\frac{1}{B(p,q)}F^{p-1}f(1-F)^{q-1},
\label{GBPDFgen}
\end{equation}
which produces an alternative form of GB PDF,
\begin{equation}
f_{GB}(x;\alpha,\beta _1,\beta _2, p,q)=\frac{\alpha}{x B(p,q)}  \left(\left(\frac{x}{\beta _1}\right)^{\alpha }+\left(\frac{x}{\beta _2}\right)^{\alpha }\right)^p \left(1-\left(\frac{x}{\beta _1}\right)^{\alpha }\right)^{q-1} \left(1+\left(\frac{x}{\beta _2}\right)^{\alpha }\right)^{-p-q},
\label{GBPDF2}
\end{equation}
equivalent to (\ref{GBPDF}). In particular, it should be mentioned that for $\alpha=1$ the seed distribution (\ref{Fseed}) contains only scale parameters $\beta_1$ and $\beta_2$, so the effect of generating B distribution via (\ref{GBCDF}) and (\ref{GBPDF2}) is the attainment of the shape parameters $p$ and $q$.

In what follows, we will be specifically interested in the $\beta_2\ll\beta_1$ circumstance since for $\beta_2 \ll x \ll \beta_1$ GB exhibits a power-law dependence,
\begin{equation}
f_{GB} \propto \left(\frac{x}{\beta _2}\right)^{-\alpha q - 1}, \hspace{1cm} 1-F_{GB} \propto \left(\frac{x}{\beta _2}\right)^{-\alpha q},
\label{GB2tail}
\end{equation}
which is also the power-law tail of GB2, but GB PDF eventually terminates at $\beta_1$, which may be the case in some of the possible negative Dragon Kings (nDK) phenomena \cite{pisarenko2012robust}, such as realized market volatility, which we discuss in Sec. \ref{Volatility}. 
  
 \section{Generalized Beta Family as Steady-State Distributions Stochastic Differential Equation\label{SDE}}
Invoking now an SDE approach to the GB family of distributions, we point out that -- with the exception of GB (and B) themselves -- all members of GB family of distributions where obtained as steady-state distributions of SDE by Hertzler \cite{hertzler2003classical}. We, however, provide our own, considerably simplified, version of those results and, in addition, obtain the SDE yielding mGB. 

First, to underscore the significance of the change of variable $y=x^\alpha$ in generalizing from B to GB family of distributions that was mentioned in Sec. \ref{GB}, we first consider the following SDE, which combines the multiplicative \cite{nelson1990arch, praetz1972distribution,fuentes2009universal,ma2014model} and Heston (Cox-Ingersoll-Ross)\cite{cox1985theory,heston1993closed} models of stochastic volatility \cite{dashti2021combined}:
 \begin{equation}
\mathrm{d}y= -\gamma(y - \theta)\mathrm{d}t + \sqrt{ \kappa^2 y+\kappa_2^2 y^2}\mathrm{d}W_t,
\label{B2SDE}
\end{equation}
where $\mathrm{d}W_t$ is the Wiener process, $\mathrm{d}W_t \sim \mathrm{N(}0,\, \mathrm{d}t \mathrm{)}$. Its steady-state distribution is a modified B2 \cite{dashti2021combined}, which can be easily derived using the standard Fokker-Planck formalism \cite{risken1996fokker,jacobs2010stochastic}:
\begin{equation}
f_{mB2}(x; \beta_2,p,q)=\frac{(1+\frac{x}{\beta_2})^{-p-q-1}(\frac{x}{\beta_2})^{-1+p}}{\beta_2 B(p,q+1)}=\frac{(p+q)(1+\frac{x}{\beta_2})^{-p-q-1}(\frac{x}{\beta_2})^{-1+p}}{q\beta_2 B(p,q)},
\label{mB2PDF}
\end{equation}
where
\begin{equation}
\label{B2beta2}
\beta_2=\frac{\kappa^2}{\kappa_2^2},
\end{equation}
\begin{equation}
p=\frac{2 \gamma \theta}{\kappa^2},
\label{mB2p}
\end{equation}
and
\begin{equation}
q=\frac{2 \gamma}{\kappa_2^2}.
\label{mB2q}
\end{equation}
This PDF is normalizable for $q > 0$ and $p>0$.\footnote{\label{SDEcorrespondence} Notice, that in \cite{dashti2021combined} we used the standard B2 PDF, \begin{equation}
f_{B2}(y; \beta_2,p,q)=\frac{(1+\frac{y}{\beta_2})^{-p-q}(\frac{y}{\beta_2})^{-1+p}}{\beta_2 B(p,q)},
\label{B2PDF}
\end{equation} 
whence $q=1+\frac{2 \gamma}{\kappa_2^2}$. The reason behind this ambiguity in the definition of $q$ stems from the fact that $p$ and $q$ are independent at the B2/GB2 level in the steady-state PDF of the respective SDE, which is not the case for B/GB as will be shown below. Similar ambiguity with respect to the value of $q$ exists for B1/GB1 for the same reason as for B2/GB2 -- that $p$ and $q$ are independent at that level and, consequently,  B1/GB1 PDF can be written either in standard or modified version.}

If we now consider the same mean-reverting process (\ref{B2SDE}) for $y=x^\alpha$, and define $\gamma^\prime=\gamma/\alpha$, $k^\prime=k/\alpha$ and $k_2^\prime=k_2/\alpha$, we obtain a stochastic process given by
\begin{equation}
\mathrm{d}x = -\gamma^\prime(x - \theta x^{1-\alpha})\mathrm{d}t + \sqrt{\kappa^{\prime 2} x^ {2-\alpha}+\kappa_2^{\prime 2} x^2 }\mathrm{d}W_t,
\label{GB2SDE}
\end{equation}
whose steady-state distribution is a mGB2, given by (compare with GB2 in \cite{dashti2021combined, dashti2021realized, dashti2020stochastic}) 
\begin{equation}
f_{mGB2}(x; \alpha, \beta_2, p,q)=\frac{\alpha (p+q)(1+({\frac{x}{\beta_2}})^{\alpha})^{-p-q-1}(\frac{x}{\beta_2})^{-1+p\alpha}}{q\beta_2 B(p,q)},
\label{GB2PDF}
\end{equation}
where
\begin{equation}
\label{GB2beta}
\beta_2{}^{\alpha}=\frac{\kappa^{\prime2}}{\kappa_2^{\prime 2}},
\end{equation}
\begin{equation}
p \hspace{.05cm} \alpha=-1+\alpha +\frac{2 \gamma^\prime \theta}{\kappa^{\prime 2}},
\label{mGB2p}
\end{equation}
and
\begin{equation}
q   \hspace{.05cm} =-1+\alpha+\frac{2 \gamma^\prime}{\kappa_2^{\prime 2}}.
\label{mGB2q}
\end{equation}
Notice the obvious renormalization of $p$ and $q$ in going from mB2 to mGB2 (or B2 to GB2 \cite{dashti2021combined}), unlike undergoing the change of variable in accordance with (\ref{B-GB}). Nonetheless, this analysis confirms that, for simplicity, it is sufficient to obtain an SDE for mB, given that the SDE for mGB will then automatically follow via aforementioned change of variable $x \to x^{\alpha}$. Analytical renormalization of $p$ and $q$ is irrelevant for numerical analysis, since they are obtained from fitting. 

The SDE formalism also makes analysis of limiting cases physically appealing. For instance, at the current GB2 level of hierarchy, substituting $\kappa=0$ directly into (\ref{GB2SDE}) yields a GIGa (IGa for $\alpha=1$ \cite{bouchaud2000wealth}) steady-state distribution \cite{ma2013distribution,ma2014model,dashti2021combined}, while substituting $\kappa_2=0$ yields GGa (Ga for $\alpha=1$ \cite{dragulescu2002probability}). (\ref{mGB2p}) and (\ref{mGB2q}) then also explain that those two cases correspond to $p \to \infty$ and $q \to \infty$ \cite{mcdonald1984some, mcdonald1995generalization, hertzler2003classical, chotikapanich2008modelling} respectively, when starting from (\ref{GB2PDF}). However, the explicit form of the steady-state distributions indicates that this just corresponds to the exponential-like decay for small values of the variable in the former case and for large values of the variable in the latter.

A generalization to an SDE for the modified B (mB) can be written as
\begin{equation}
\mathrm{d}x= -\gamma(x - \theta)\mathrm{d}t + \sqrt{\kappa^2 x \left( 1-\frac{\kappa_1^2}{\kappa^2} x \right) \left(1 +\frac{\kappa_2^2}{\kappa^2} x\right)}\mathrm{d}W_t,
\label{mBSDE}
\end{equation}
or using
\begin{equation}
\beta_{1,2}=\frac{\kappa^{2}}{\kappa_{1,2}^{2}},
\label{Bbeta}
\end{equation}
and rescaling, an alternative, simplified form of the SDE producing mB can be written as 
\begin{equation}
\mathrm{d}x= -\gamma(x - \theta)\mathrm{d}t + \sqrt{ x \left( 1-\frac{x}{\beta_1} \right) \left(1 +\frac{x}{\beta_2} \right)}\mathrm{d}W_t.
\label{mBSDE2}
\end{equation}
But, as before, the SDE given by (\ref{mBSDE}) allows us to easily trace B-hierarchy: $\kappa_1=0$ yields the B2 family discussed above, with further $\kappa=0$ and $\kappa_2=0$ yielding IGa and Ga respectively; conversely, $\kappa_2=0$ yields the B1 family, with further $\kappa_1=0$ yielding Ga. As discussed above, this can be immediately upgraded to the GB hierarchy via the $x \to x^{\alpha}$ change of variable, with the result summarized as follows:
\begin{equation}
\begin{split}
mGB \xrightarrow{\kappa_1=0} GB2 \xrightarrow{\kappa=0} GIGa \\
mGB \xrightarrow{\kappa_1=0} GB2 \xrightarrow{\kappa_2=0} GGa \\
mGB \xrightarrow{\kappa_2=0} GB1 \xrightarrow{\kappa_1=0} GGa
\end{split}
\label{GBhierarchy}
\end{equation}
As was mentioned before, GB2/GB1 distributions can be written either in its standard or modified form, depending on the definition of $q$. 

It is obvious form (\ref{GBhierarchy}) that (G)Ga is a tying link between (G)B2 and and (G)B1. This can be easily seen by considering the following SDE:
\begin{equation}
\mathrm{d}x= -\gamma(x - \theta)\mathrm{d}t + \sqrt{ \left( \kappa^2 x+(2c-1) \tilde{\kappa}^2 x \right)}\mathrm{d}W_t,
\label{B2B1SDE}
\end{equation}
where $0 \le c \le 1$. For $c=0$ we have $\tilde{\kappa}=\kappa_1$ and B1 above; for $c=1/2$ we have G; for $c=1$ we have $\tilde{\kappa}=\kappa_2$ and B2 above. This simply means that as $c$ changes we observe a transition from a PDF defined on a finite interval, (G)B, to a PDF with a power-law tail, (G)B2, via a distribution with an exponential-like tail. 

The PDF and the CDF of the steady-steady distribution obtained from (\ref{mBSDE2}) are given respectively by
\begin{equation}
f_{mB}(x;\beta _1,\beta _2, p,q)=\frac{(p+q)\left(1+\frac{\beta _1}{\beta _2}\right)^{p+1}  \left(\frac{x}{\beta _1}\right)^{p-1} \left(1-\frac{x}{\beta _1}\right)^{q-1} \left(1+\frac{x}{\beta _2}\right)^{-p-q-1}}{\beta _1 \left(p+\left(1+\frac{\beta _1}{\beta _2}\right) q\right) B(p,q)},
\label{mBPDF}
\end{equation}
and
\begin{equation}
F_{mB}(x;\beta _1,\beta _2, p,q)=I\left({\frac{\frac{x}{\beta _1}+\frac{x}{\beta _2}}{1+\frac{x}{\beta _2}}};p,q\right) + \frac{1}{B(p,q) \left(q+\left(\frac{\beta _2}{\beta _1}\right) (p+q)\right)}  \left(\frac{1-\frac{x}{\beta _1}}{1+\frac{x}{\beta _2}}\right)^q\left(\frac{\left(1+\frac{\beta _1}{\beta _2}\right) \frac{x}{\beta _1}}{\left(1+\frac{x}{\beta _2}\right)}\right)^p,
\label{mBCDF}
\end{equation}
where
\begin{equation}
p=2 \gamma \theta,
\label{mBp}
\end{equation}
and
\begin{equation}
q=2 \gamma(\beta_1-\theta)\left(1+\frac{\beta_1}{\beta_2}\right)^{-1}.
\label{mBq}
\end{equation}
One version of the mGB PDF and the CDF related to GB can be obtained via the change of variable (\ref{B-GB}) applied to (\ref{mBPDF}) and (\ref{mBCDF}) and are given respectively by
\begin{equation}
f_{mGB}(x;\alpha,\beta _1,\beta _2, p,q)=\frac{\alpha(p+q)\left(1+\left(\frac{\beta _1}{\beta _2}\right)^\alpha\right)^{p+1}  \left(\frac{x}{\beta _1}\right)^{\alpha p-1} \left(1-\left(\frac{x}{\beta _1}\right)^\alpha\right)^{q-1} \left(1+\left(\frac{x}{\beta _2}\right)^\alpha\right)^{-p-q-1}}{\beta _1 \left(p+\left(1+\left(\frac{\beta _1}{\beta _2}\right)^\alpha\right) q\right) B(p,q)},
\label{mGBPDF}
\end{equation}
and
\begin{equation}
\resizebox{1.0\hsize}{!}{$
F_{mGB}(x;\alpha,\beta _1,\beta _2, p,q)=I\left({\frac{\left(\frac{x}{\beta _1}\right)^\alpha+\left(\frac{x}{\beta _2}\right)^\alpha}{1+\left(\frac{x}{\beta _2}\right)^\alpha}};p,q\right) + \frac{1}{B(p,q) \left(q+\left(\frac{\beta _2}{\beta _1}\right)^{\alpha } (p+q)\right)} \left(\frac{1-\left(\frac{x}{\beta _1}\right)^\alpha}{1+\left(\frac{x}{\beta _2}\right)^\alpha}\right)^q\left(\frac{\left(1+\left(\frac{\beta _1}{\beta _2}\right)^\alpha\right) \left(\frac{x}{\beta _1}\right)^\alpha}{\left(1+\left(\frac{x}{\beta _2}\right)^\alpha\right)}\right)^p,
$}
\label{mGBCDF}
\end{equation}
and also
\begin{equation}
\resizebox{1.0\hsize}{!}{$
1-F_{mGB}(x;\alpha,\beta _1,\beta _2, p,q)=I\left({\frac{1-\left(\frac{x}{\beta _1}\right)^{\alpha }}{1+\left(\frac{x}{\beta _2}\right)^{\alpha }}};q,p\right) -\frac{1}{B(p,q) \left(q+\left(\frac{\beta _2}{\beta _1}\right)^{\alpha } (p+q)\right)} \left(\frac{1-\left(\frac{x}{\beta _1}\right)^\alpha}{1+\left(\frac{x}{\beta _2}\right)^\alpha}\right)^q\left(\frac{\left(1+\left(\frac{\beta _1}{\beta _2}\right)^\alpha\right) \left(\frac{x}{\beta _1}\right)^\alpha}{\left(1+\left(\frac{x}{\beta _2}\right)^\alpha\right)}\right)^p.
$}
\label{mGBCDFtail}
\end{equation}
The second term in (\ref{mGBCDF}) is the difference between $F_{mGB}$ and $F_{GB}$ in (\ref{GBCDF}) and is zero at zero and at $\beta_1$. Due to this difference, the asymptotic behaviors of $F_{GB}$ and $F_{mGB}$ on approach to $\beta_1$, $x\rightarrow \beta_1$, develop a considerable contrast in the limit $\beta_2 \ll \beta_1$ of interest here: 

\begin{equation}
1-F_{GB} \approx \frac{1}{qB(p,q)} \left(\frac{1-\left(\frac{x}{\beta _1}\right)^\alpha}{1+\left(\frac{x}{\beta _2}\right)^\alpha}\right)^q \approx \frac{1}{qB(p,q)} \left(\frac{1-\left(\frac{x}{\beta _1}\right)^\alpha}{1+\left(\frac{\beta _1}{\beta _2}\right)^\alpha}\right)^q
\label{GBCDFbeta1}
\end{equation}

\begin{equation}
1-F_{mGB} \approx \frac{1+\frac{p}{q}}{qB(p,q)} \left(\frac{1-\left(\frac{x}{\beta _1}\right)^\alpha}{1+\left(\frac{x}{\beta _2}\right)^\alpha}\right)^q \left(\frac{\beta_2}{\beta_1} \right)^{\alpha} \approx \frac{1+\frac{p}{q}}{qB(p,q)} \left(\frac{1-\left(\frac{x}{\beta _1}\right)^\alpha}{1+\left(\frac{\beta _1}{\beta _2}\right)^\alpha}\right)^q \left(\frac{\beta_2}{\beta_1} \right)^{\alpha}
\label{mGBCDFbeta1}
\end{equation}
that is $1-F_{mGB}$ drops off to zero ($F_{mGB}$ saturates to unity) faster than $1-F_{GB}$ due to the factor $\left(\frac{\beta _2}{\beta _1}\right)^\alpha$.

Another mGB distribution is obtained from the SDE for B with $x$ replaced by $x^\alpha$:
\begin{equation}
\mathrm{d}x= -\gamma(x - \theta x^{1-\alpha})\mathrm{d}t + \sqrt{ x^{2-\alpha} \left( 1-\left(\frac{x}{\beta_1}\right)^\alpha \right) \left(1 +\left(\frac{x}{\beta_2}\right)^\alpha \right)}\mathrm{d}W_t,
\label{mGBSDE}
\end{equation}
PDF and CDF of the steady-state solution are given respectively by
\begin{equation}
\tilde{f}_{mGB}(x;\alpha,\beta _1,\beta _2, p,q)=\frac{\alpha\left(\frac{x}{\beta _1}\right)^{\alpha  p-1} \left(1-\left(\frac{x}{\beta _1}\right)^{\alpha }\right)^{q-\frac{1}{\alpha }} \left(1+\left(\frac{x}{\beta _2}\right)^{\alpha }\right)^{-p-q-1}}{\beta _1 B \left(p,q-\frac{1}{\alpha }+1\right) \, _2F_1\left(p,p+q+1;p+q-\frac{1}{\alpha }+1;-\left(\frac{\beta _1}{\beta _2}\right)^{\alpha }\right)},
\label{mGBPDF2}
\end{equation}
and
\begin{equation}
\resizebox{1.0\hsize}{!}{$
\begin{multlined}
\tilde{F}_{mGB}(x;\alpha,\beta _1,\beta _2, p,q)= \\
\frac{\left(\frac{x}{\beta_1}\right)^{\alpha p} \left(\frac{1}{p}F_1\left(p;\frac{1}{\alpha }-q,p+q;p+1;\left(\frac{x}{\beta_1}\right)^{\alpha},- \left(\frac{x}{\beta _2}\right)^{\alpha }\right)-\frac{1}{p+1} \left(\frac{x}{\beta _2}\right)^{\alpha } F_1\left(p+1;\frac{1}{\alpha }-q,p+q+1;p+2;\left(\frac{x}{\beta_1}\right)^{\alpha},- \left(\frac{x}{\beta _2}\right)^{\alpha }\right)\right)}{ B \left(p,q-\frac{1}{\alpha }+1\right) \, _2F_1\left(p,p+q+1;p+q-\frac{1}{\alpha }+1;-\left(\frac{\beta _1}{\beta _2}\right)^{\alpha }\right)},
\end{multlined}
$}
\label{mGBCDF2}
\end{equation}
where 
\begin{equation}
p \alpha = \alpha -1 + 2 \gamma  \theta,
\end{equation}
\label{mGBp}
\begin{equation}
q \alpha =1-\alpha +2\gamma(\beta_1-\theta)\left(1+\frac{\beta_1}{\beta_2}\right)^{-1},
\label{mGBq}
\end{equation}
and $ \, _2F_1$ and $F_1$ are hypergeometric and Appell hypergeometric functions respectively \cite{nist2022digital}.

 \section{Market Volatility as Possible Example of Application of Generalized Beta Distribution\label{Volatility}}
 
Realized volatility $RV$ is the square root of realized variance, which is defined as follows
\begin{equation}
RV^2=100^2\times\frac{252}{n}\sum_{i=1}^nr_i^2
\label{RV2}
\end{equation}
where
\begin{equation}
r_i=\ln\frac{S_{i}}{S_{i-1}}
\label{ri}
\end{equation}
are daily returns and $S_{i}$ is the reference (closing) price on day $i$. This is an annualized value, where $252$ represents the number of trading days in a year. In particular, $n=1$ are the daily returns and $n=21$ are monthly returns (typical number of trading days in a month). Here we present results for $n=1, 2, 3, 5, 7, 9, 13, 17, 21$.
 
We fitted distributions of RV for S\&P index from 1970 through May of 2021, which covers four major financial upheavals: Savings and Loan Crisis, Tech Bubble, Great Recession, and Covid Pandemic. In a companion manuscript \cite{liu2022dragon}, we undertake a far more detailed numerical analysis, which includes actual time-series of RV identifying the far-end-tail points of the distribution, linear tail fitting of RV distribution on a log-log scale meant to test conformity to power-law behavior, and p-value test for DK \cite{pisarenko2012robust}. Here, we limit our discussion to fitting RV distribution using (\ref{GBPDF})-(\ref{GBCDF}) and (\ref{mGBPDF})-(\ref{mGBCDF}), including the confidence intervals (CI) of these fits \cite{janczura2012black}.

Fitting was conducted using Bayesian for PDF and Gradient Descendant for CDF and we find very small differences in the parameters of the distribution between the two techniques -- at most 8\% for GB (\ref{GBPDF})-(\ref{GBCDF}) and at most 3\% for mGB (\ref{mGBPDF})-(\ref{mGBCDF}). Additionally, the two-sample KS statistic between (\ref{mGBCDF}) and (\ref{mGBCDF2}) is $0.0015 \ll 0.0169$, the latter being the standard value for the number of points in our sample, as per \cite{knuth1998art} with alpha value being $0.05$. Given such negligible difference, relative simplicity of (\ref{mGBPDF})-(\ref{mGBCDF}) vis-a-vis the actual solution (\ref{mGBPDF2})-(\ref{mGBCDF2}) of the SDE (\ref{mGBSDE}), qualifies the former as a near exact solution of the SDE and explains its choice for fitting.

\begin{figure}[tb]
	\centering
		\includegraphics[width = 1. \textwidth]{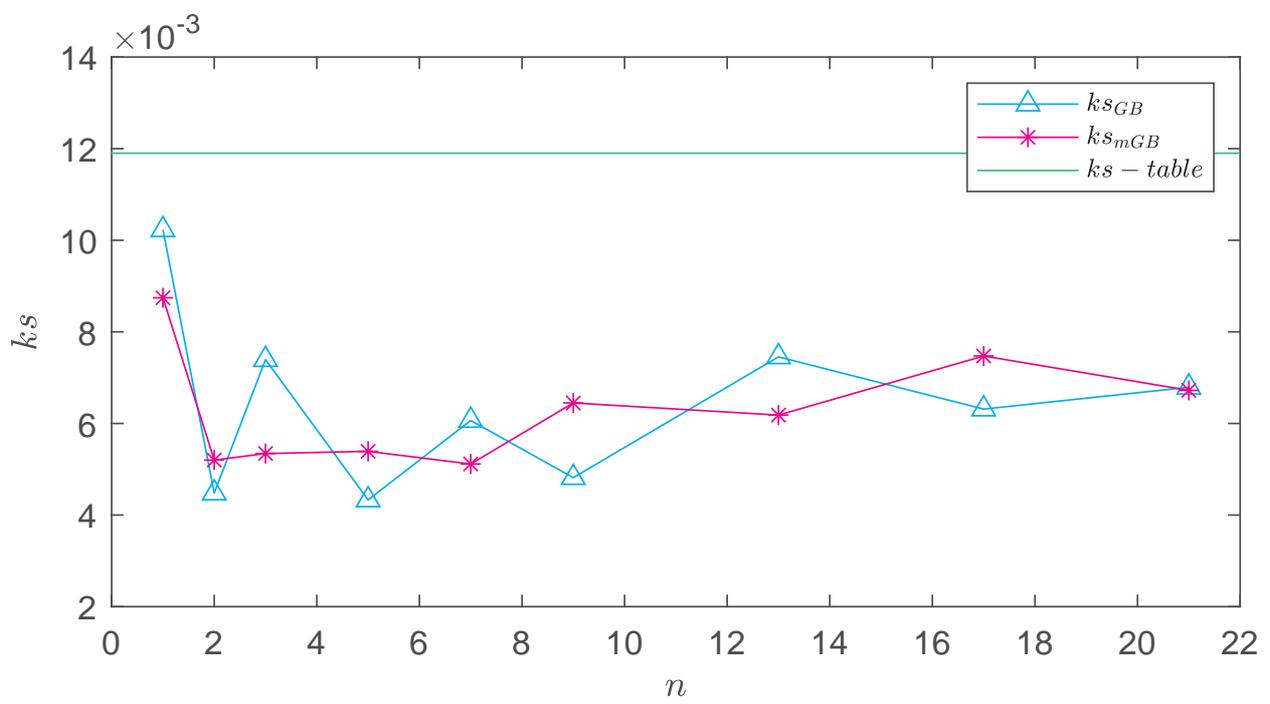}
	\caption{$ks$ vs $n$ for GB and mGB fits; horizontal line shows the minimal value of goodness-of-fit $ks$ from Tables \ref{parGB} and \ref{parmGB}.}
\label{KSn}
\end{figure}

\begin{figure}[htbp!]
	\centering
		\includegraphics[width = 1. \textwidth]{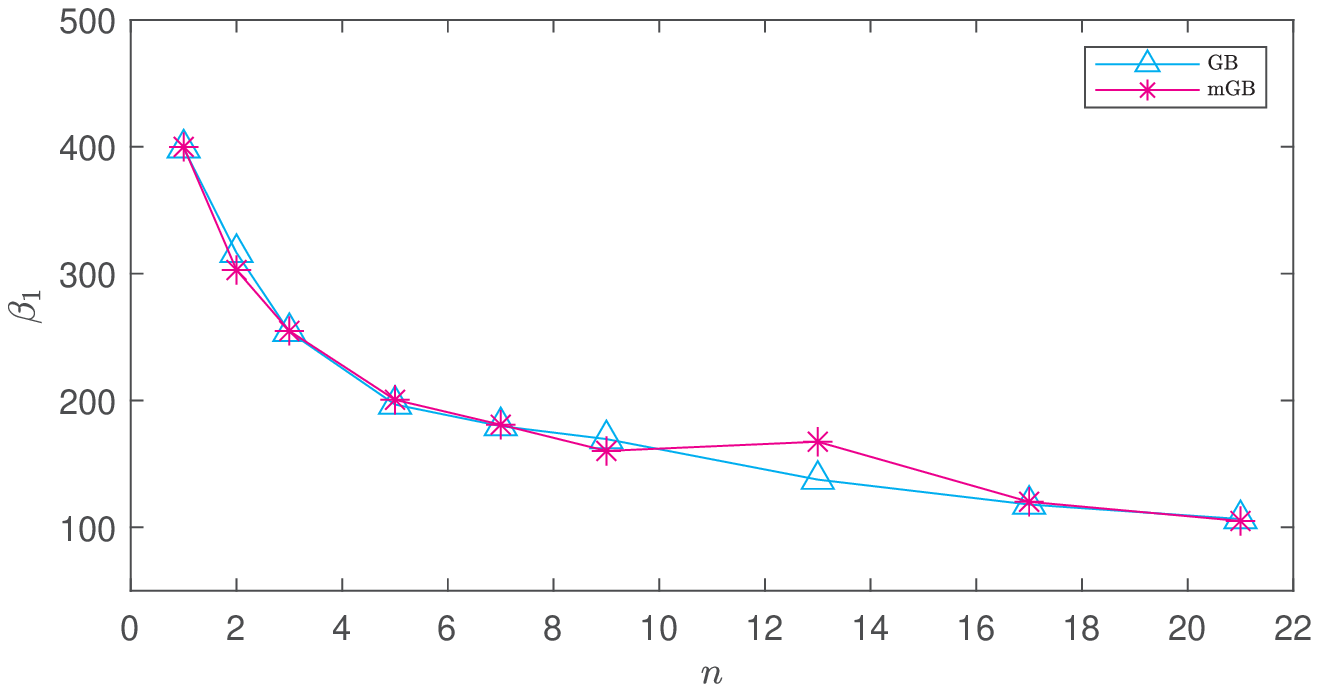}\\
		\includegraphics[width = 1. \textwidth]{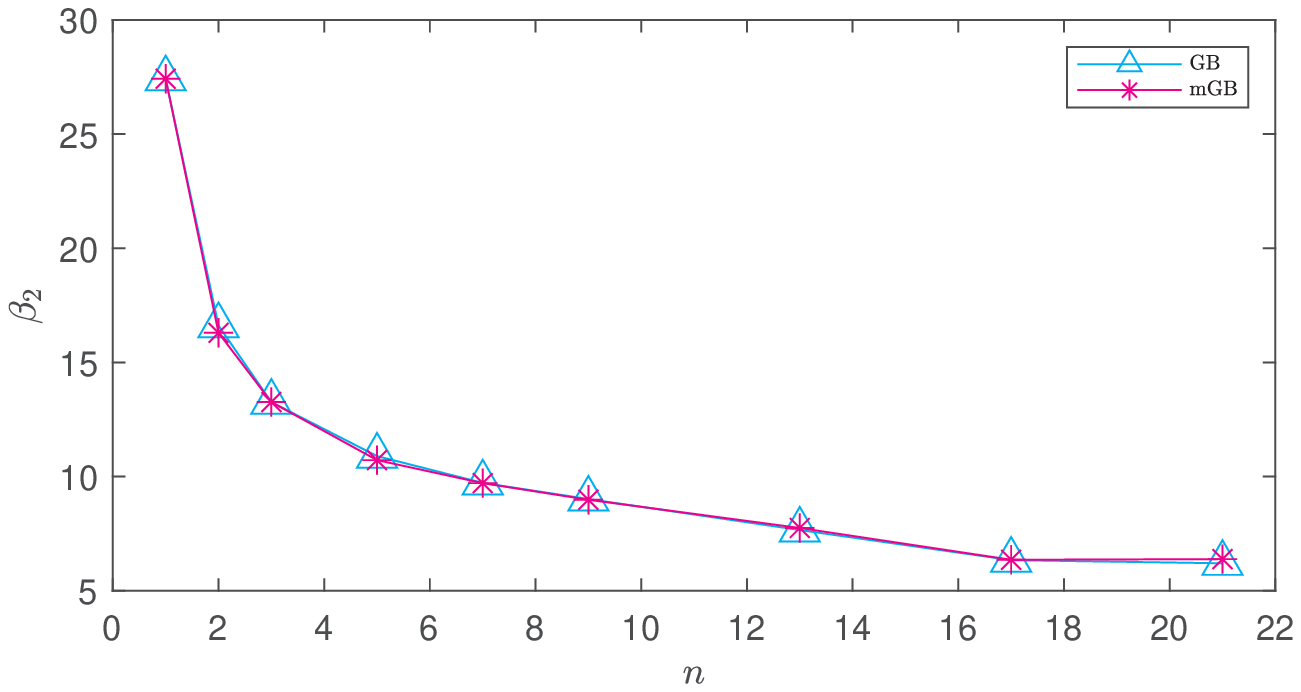}
	\caption{$\beta_{1}$ and $\beta_{2}$ vs $n$ for GB and mGB fits.}
\label{beta1beta2n}
\end{figure}

Tables \ref{parGB} and \ref{parmGB} list the parameters of GB and mGB distributions obtained from fitting the RV distribution, the corresponding Kolmogorov-Smirnov (KS) statistic, and the goodness-of-fit KS values for our RV sample size, as per table in \cite{massey1985kolmogorov} with alpha value being $0.05$. Figs. \ref{KSn} and \ref{beta1beta2n} show plots of KS statistic and scale parameters $\beta_1$ and $\beta_2$ from the tables as a function of $n$. Figs. \ref{n1}$-$\ref{n21} show GB and mGB fits of RV data -- 1-CDF, or complimentary CDF (ccdf) -- with the corresponding 95\% CI, on a log-log scale. 

One noteworthy feature in those plots is that a relatively small subset of data points moves up from the straight line (on a log-log scale) portion of the GB and mGB tails, (\ref{GB2tail}), and outside their CI, which can be viewed as "potential DK." This feature is further analyzed in \cite{liu2022dragon} using the p-value test \cite{pisarenko2012robust}. However, the data invariably falls back in and indicates termination at final values, as intended to be explained by GB and mGB. 

\begin{table}[!htb]
\caption{Parameters of GB fit of RV distribution, its KS statistic $ks$, and goodness-of-fit value of KS for our RV sample sizes.}
\label{parGB}
\begin{center}
\begin{tabular}{ c c c c} 
\hline
           n &  $ GB(x;\alpha,\beta_{1},\beta_{2},p,q) $  &          $ks$&         $ks$-table \\
\hline
1 &	(1.5457, 398.8160, 27.4217, 0.6648, 2.7871) &  0.0102&0.0119\\
\hline
2 &	(2.0163, 316.3938, 16.6113, 0.8805, 1.8097) & 0.0044&0.0119\\
\hline
3 & 	(2.1444, 254.1085, 13.2608, 1.2549, 1.6824) & 0.0074&0.0119\\
\hline
5 & 	(2.2971, 196.7883, 10.8962, 1.7834, 1.5348)& 0.0043&0.0119\\
\hline
7 & 	(2.4789, 179.8124, 9.7236, 2.1369, 1.3815) & 0.0060&0.0120\\
\hline
9 &	 (2.4734, 169.5618, 9.0164, 2.5880, 1.3855) & 0.0048&0.0120\\
\hline
13 & (2.4317, 137.6122, 7.6590, 3.8712, 1.4172) & 0.0075&0.0120\\
\hline
17 & (2.2842, 117.9511, 6.3396, 6.1014, 1.5241) &	0.0063&0.0120\\
\hline
21 & (2.3979, 106.5157, 6.2021, 6.5453, 1.4415) & 0.0068&0.0120\\
\hline
\hline
\end{tabular}
\end{center}
\end{table}

\begin{table}[!htb]
\caption{Parameters of mGB fit of RV distribution, its KS statistic $ks$, and goodness-of-fit value of KS for our RV sample sizes.}
\label{parmGB}
\begin{center}
\begin{tabular}{ c c c c} 
\hline
           n &  $ mGB(x;\alpha,\beta_{1},\beta_{2},p,q) $  &          $ks$&         $ks$-table\\
\hline
1 &	(1.5500, 399.9009, 27.4233, 0.6519, 1.7828) &  0.0087&0.0119\\
\hline
2 &	(1.9541, 302.8320, 16.2974, 0.9384, 0.8642) & 0.0052&0.0119\\
\hline
3 & 	(2.1195, 254.8331, 13.2632, 1.25611, 0.6836) & 0.0053&0.0119\\
\hline
5 & 	(2.3708, 200.5519, 10.7210, 1.7255, 0.4456)& 0.0054&0.0119\\
\hline
7 & 	(2.4744, 180.8711, 9.7136, 2.1430, 0.3848) & 0.0051&0.0120\\
\hline
9 &	 (2.5239, 160.224, 8.9839, 2.5856, 0.3582) & 0.0064&0.0120\\
\hline
13 & (2.4506, 167.4719, 7.7488, 	3.7661, 0.4092) & 0.0062&0.0120\\
\hline
17 & (2.3026, 120.1110, 6.3561, 6.1121, 0.5403) &	0.0074&0.0120\\
\hline
21 & (2.4016, 104.9925, 6.3853, 6.3429, 0.50434) & 0.0067&0.0120\\
\hline
\hline
\end{tabular}
\end{center}
\end{table}

\clearpage
\newpage

\begin{landscape}

\begin{figure}[htbp!]
\centering
\includegraphics[width = 1.2 \textwidth]{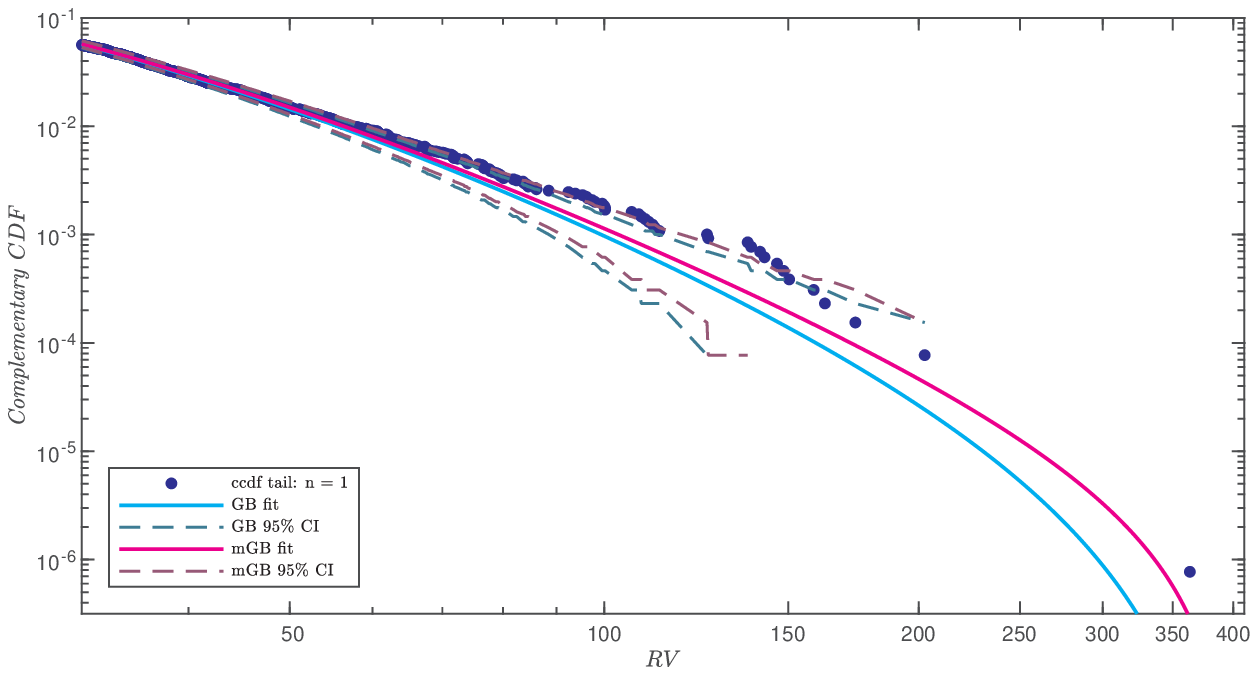}
\caption{GB and mGB fits of RV, with respective CI, for $n=1$.}
\label{n1}
\end{figure}

\begin{figure}[tb]
\centering
\includegraphics[width = 1.2 \textwidth]{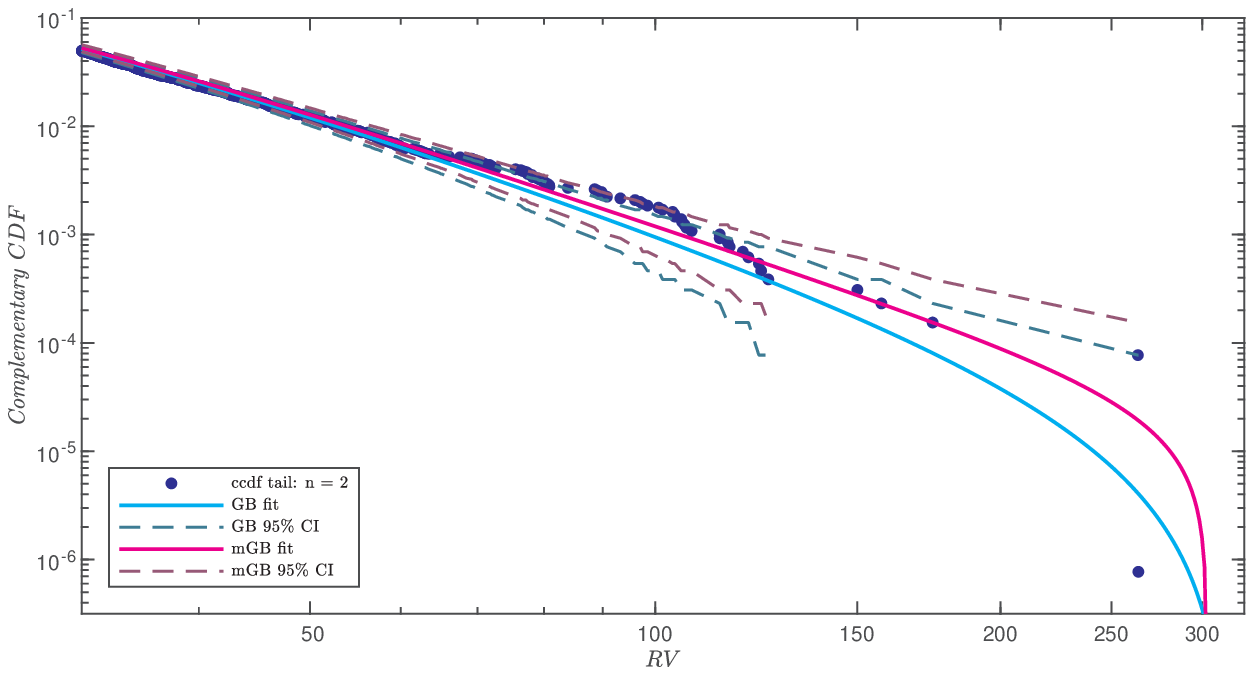}
\caption{GB and mGB fits of RV, with respective CI, for $n=2$.}
\label{n2}
\end{figure}

\begin{figure}[tb]
\centering
\includegraphics[width = 1.2 \textwidth]{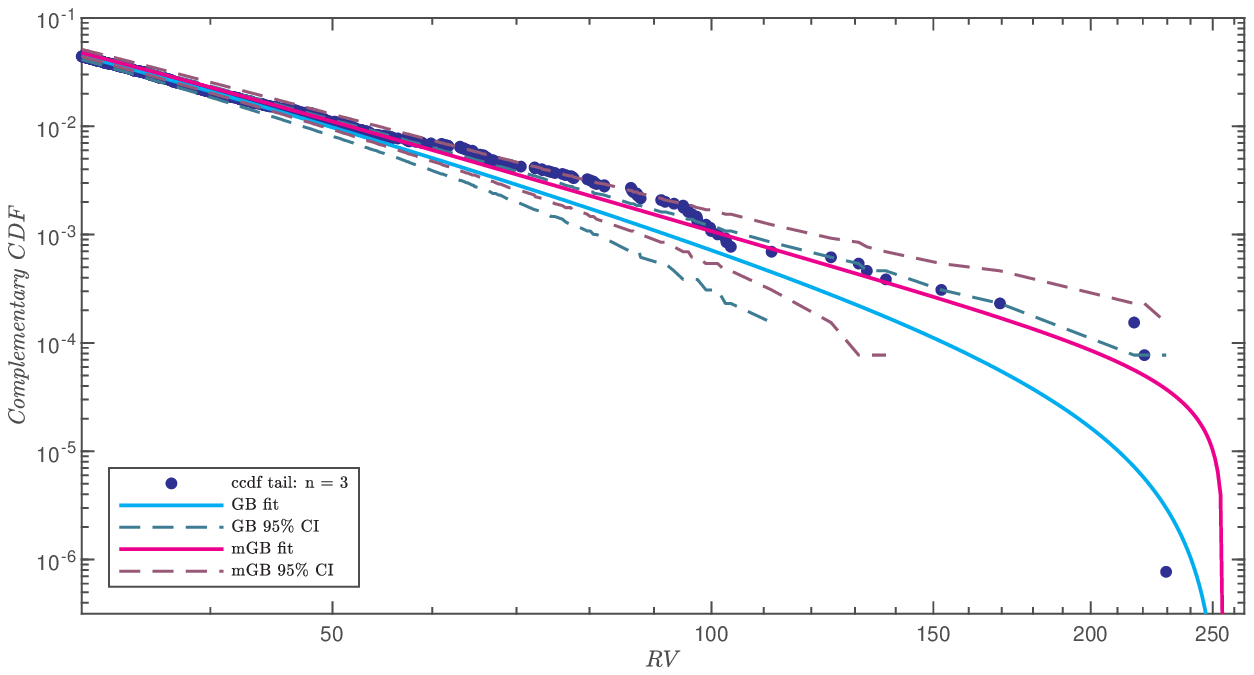}
\caption{GB and mGB fits of RV, with respective CI, for $n=3$.}
\label{n3}
\end{figure}

\begin{figure}[tb]
\centering
\includegraphics[width = 1.2 \textwidth]{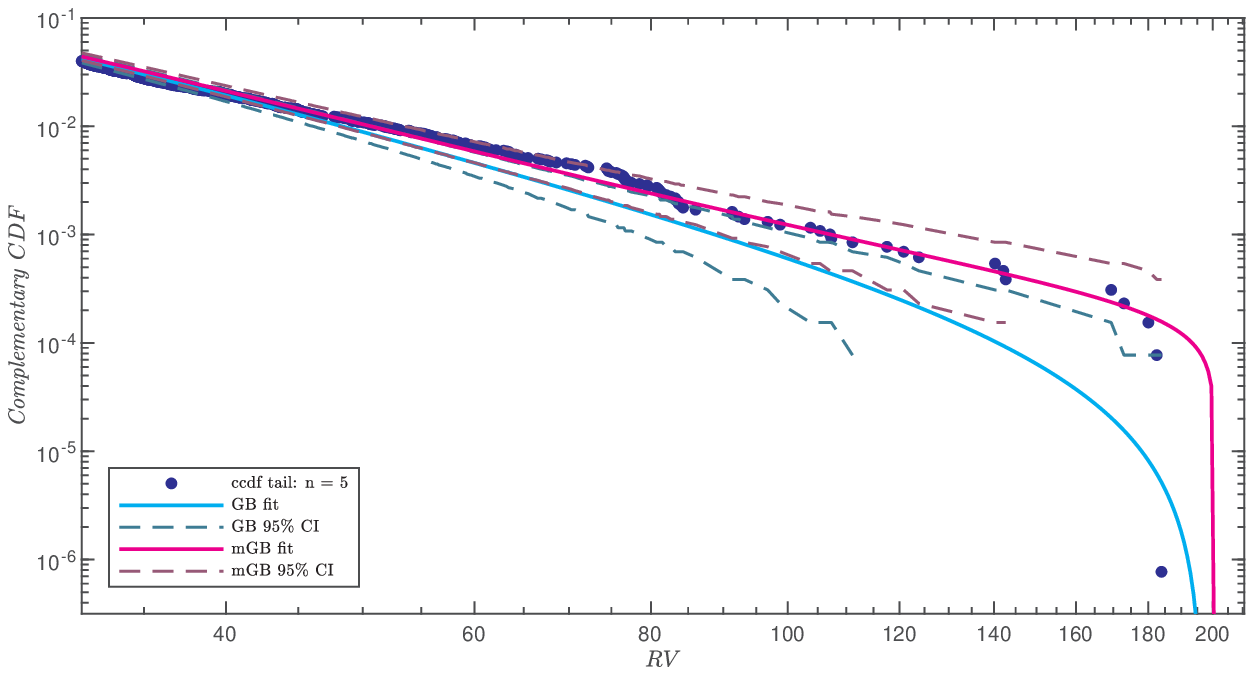}
\caption{GB and mGB fits of RV, with respective CI, for $n=5$.}
\label{n5}
\end{figure}

\begin{figure}[tb]
\centering
\includegraphics[width = 1.2 \textwidth]{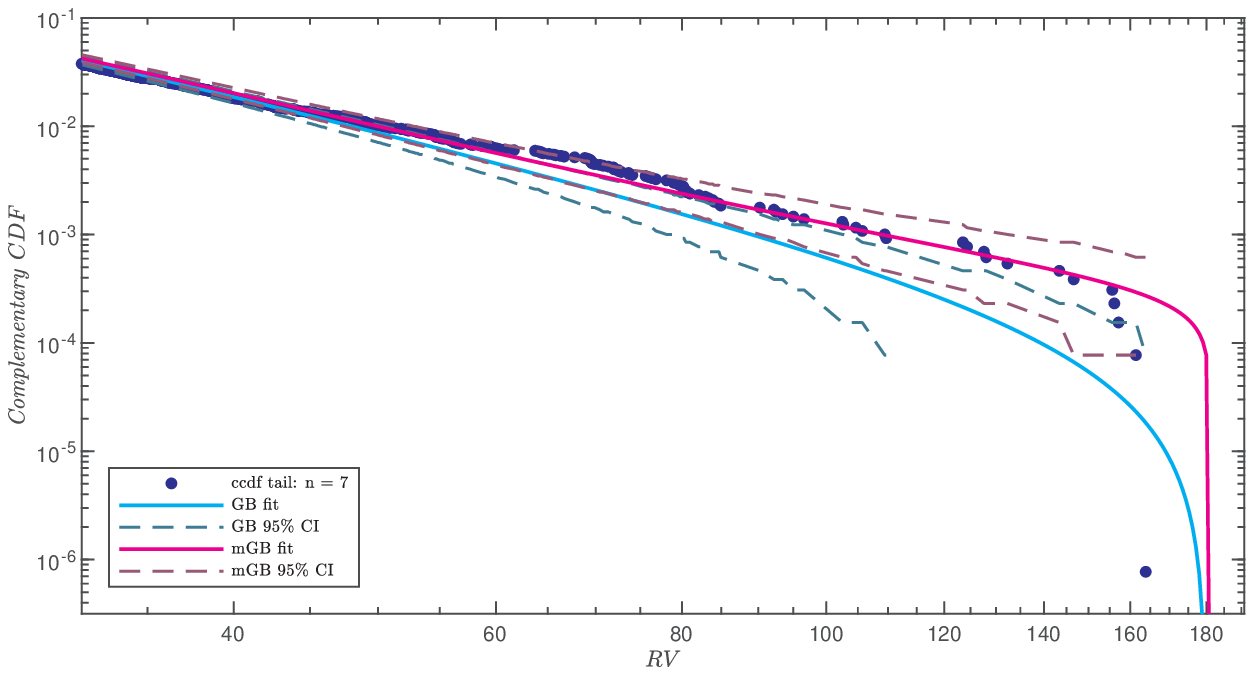}
\caption{GB and mGB fits of RV, with respective CI, for $n=7$.}
\label{n7}
\end{figure}

\begin{figure}[tb]
\centering
\includegraphics[width = 1.2 \textwidth]{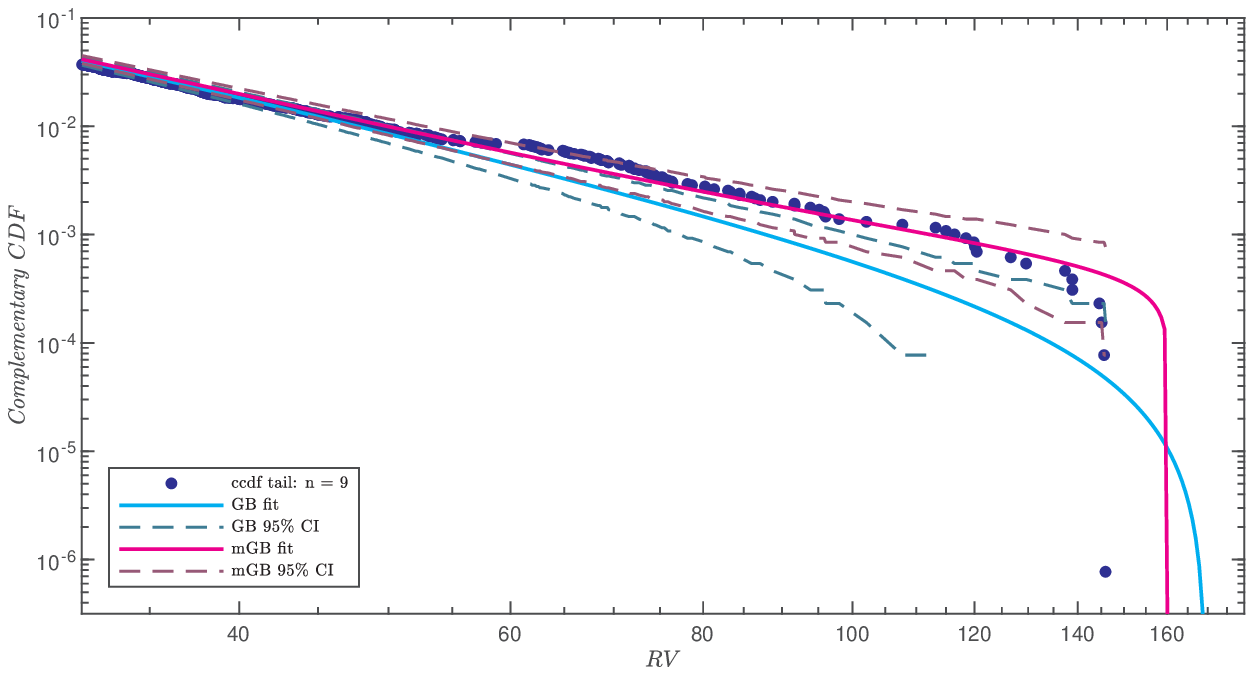}
\caption{GB and mGB fits of RV, with respective CI, for $n=9$.}
\label{n9}
\end{figure}

\begin{figure}[tb]
\centering
\includegraphics[width = 1.2 \textwidth]{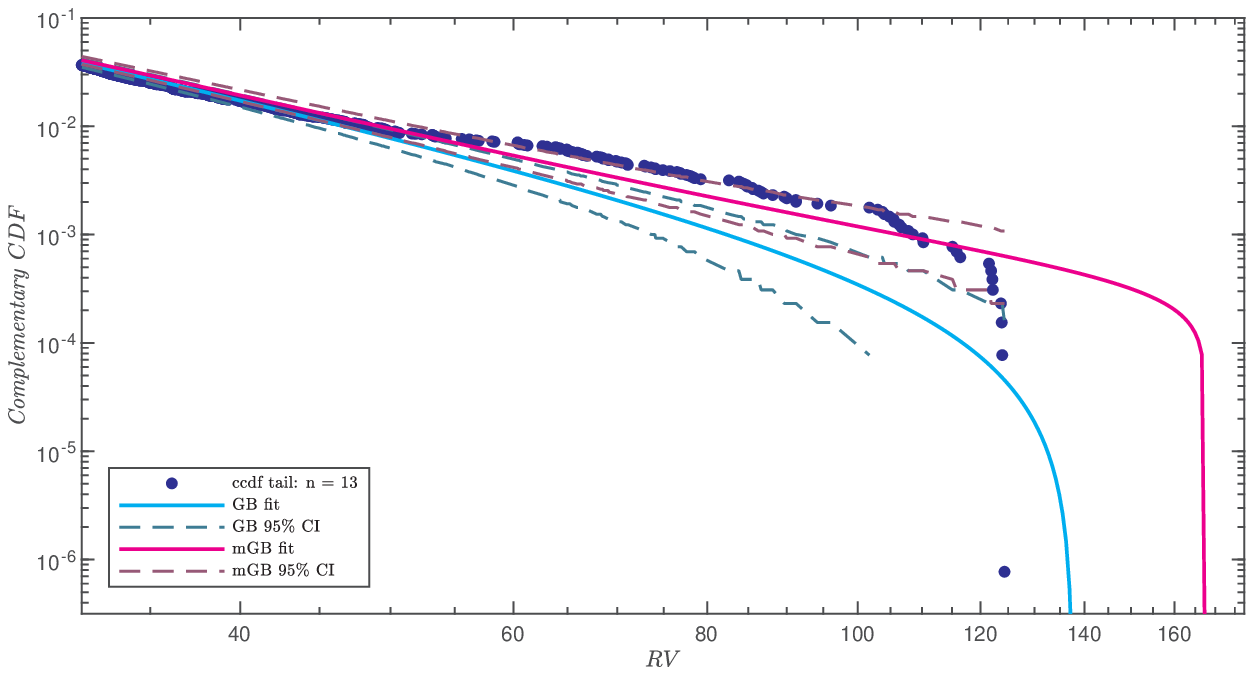}
\caption{GB and mGB fits of RV, with respective CI, for $n=13$.}
\label{n13}
\end{figure}

\begin{figure}[tb]
\centering
\includegraphics[width = 1.2 \textwidth]{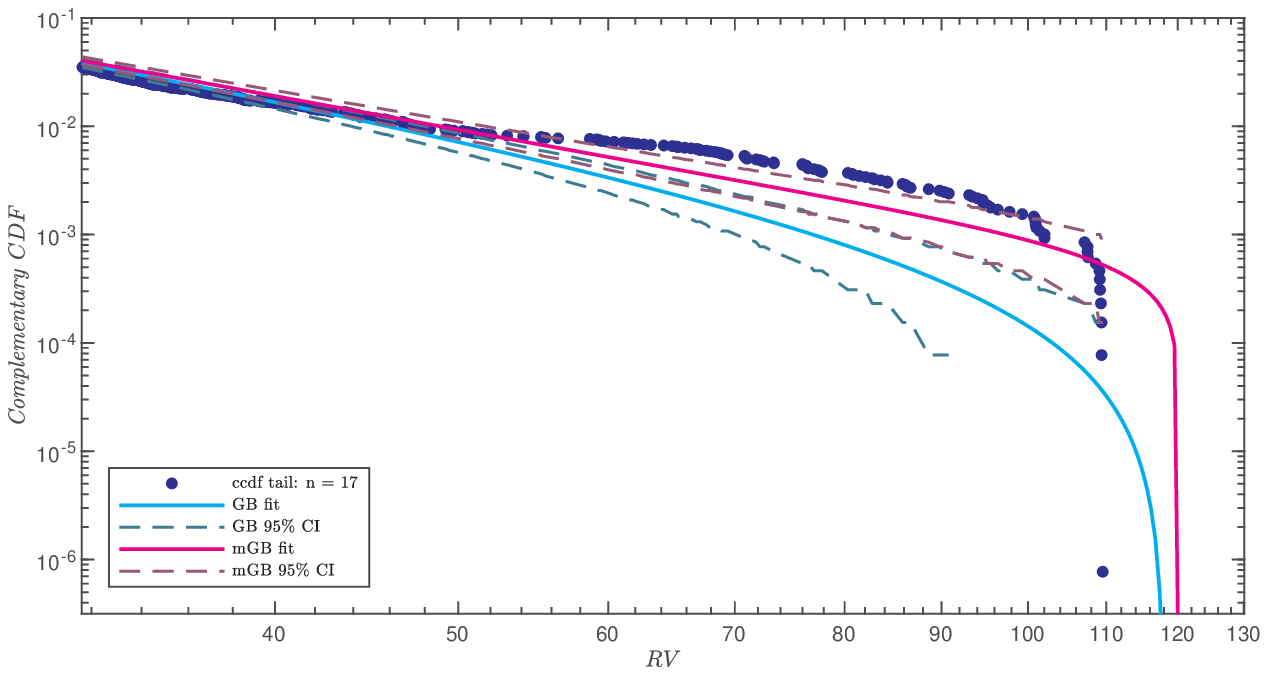}
\caption{GB and mGB fits of RV, with respective CI, for $n=17$.}
\label{n17}
\end{figure}

\begin{figure}[tb]
\centering
\includegraphics[width = 1.2 \textwidth]{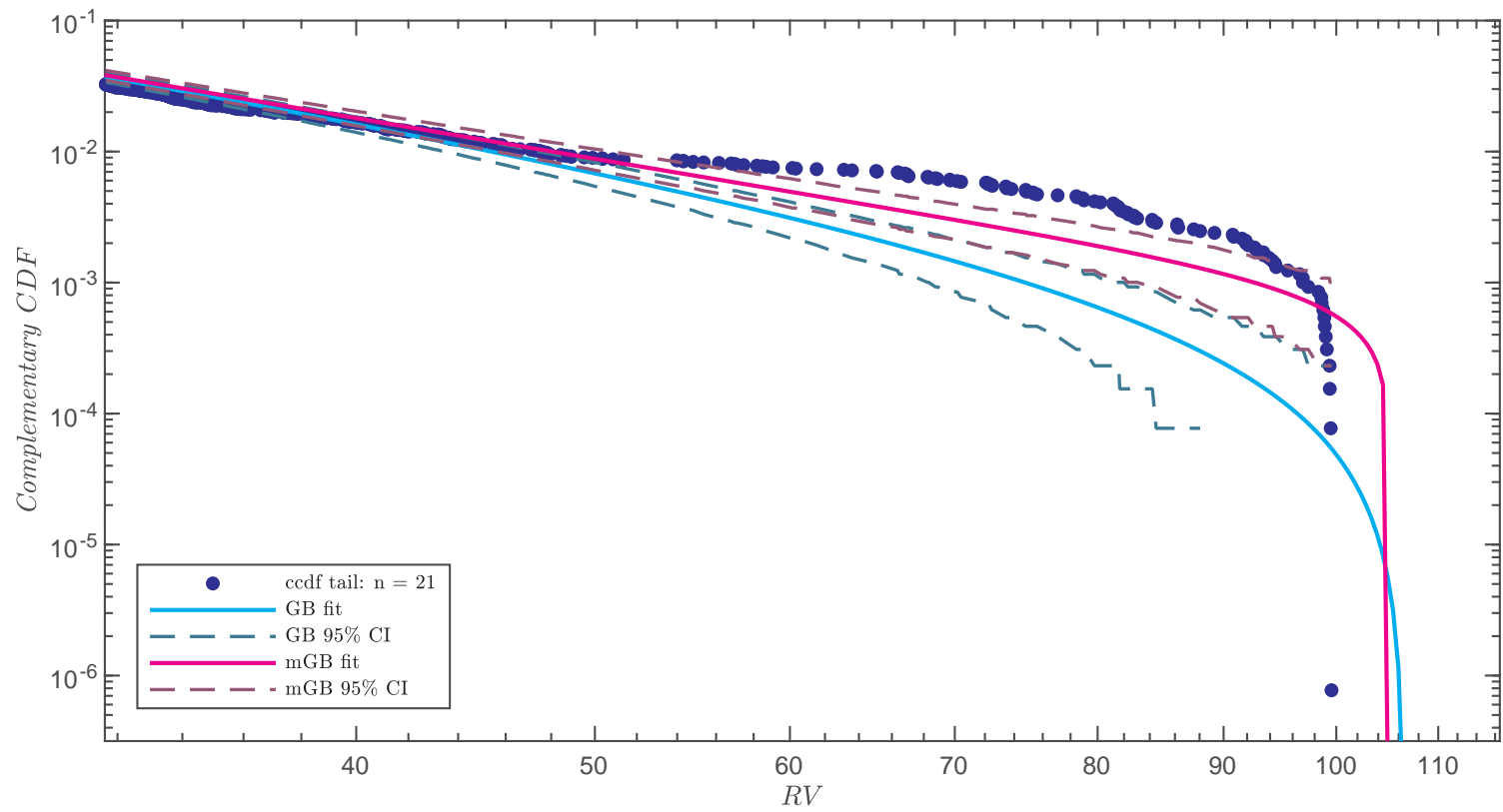}
\caption{GB and mGB fits of RV, with respective CI, for $n=21$.}
\label{n21}
\end{figure}

\end{landscape}

\section{Summary\label{Summary}}	

Main results of this article can be summarized as follows:

\begin{itemize}
  \item We introduced an alternative form of Generalized Beta PDF (\ref{GBPDF}) and evaluated its CDF (\ref{GBCDF}), the latter belonging to a class of Incomplete Beta Function generated distributions.
  \item We identified an SDE (\ref{mGBSDE}), which produces a modified Generalized Beta distribution as its steady-state solution, whose PDF and CDF are given respectively by (\ref{mGBPDF2}) and (\ref{mGBCDF2}).
  \item A similar yet simpler modified Generalized Beta distribution (\ref{mGBPDF})-(\ref{mGBCDF}) was constructed via a change of variable (\ref{B-GB}) from a modified Beta distribution (\ref{mBPDF})-(\ref{mBCDF}), which is a steady-state distribution of the $\alpha=1$ SDE, (\ref{mBSDE2}).
  \item We showed that an expanded version of (\ref{mBSDE2}), (\ref{mBSDE}) allows for a physically appealing explanation of the Generalized Beta hierarchy, (\ref{GBhierarchy}), as well as of natural link between (Generalized) Beta 1 and (Generalized) Beta 2 from (\ref{B2B1SDE}).
  \item We argued that Generalized Beta distribution is particularly useful in the circumstance when the distribution is characterized by a long power-law tail followed by a drop-off at a finite value of the variable. This has been applied to the distribution of realized volatility of the S\&P500 index, which exhibits behavior characteristic of ``Negative Dragon Kings." Specifically, we found that power-law tails, leading potentially to "Black Swan" effects, are abruptly terminated, capping realized volatility at very large but finite values. 
  
  One interesting nuance to the above observation is that a small subset of data in the tails showed upward deviation from the linear part of the tails (on the log-log scale) and outside the confidence intervals of GB and mGB fits (as well as that of linear fit, all confirmed using p-value test \cite{liu2022dragon}), which can be viewed as potential to develop "Dragon Kings." However, it is always followed by a rapid dropdown, as seen in Figs. \ref{n1} - \ref{n21}.
  
  The largest values of realized volatility of S\&P500 were, as expected, due to the most virulent market calamities: Savings and Loan crisis, Tech Bubble, Financial Crisis and Covid Pandemic. However, at least for S\&P 500 -- perhaps due to the broad spectrum of the index and the size and strength of companies it represents -- market mechanism seem to limit its price from below and thus to disallow emergence of ``Dragon Kings" or even development of "Black Swans" beyond some limiting value (which is not to state that the upper limit could not move upward in the future). To reiterate, Generalized Beta seems to be well suited for such situations and in the particular case of realized volatility provides a very good overall fit of its distribution.
\end{itemize} 

In the future, we will further examine applicability of Generalized Beta to phenomena bounded from above yet exhibiting a well-established power-law tail over the large portion of the distribution.

\section {Acknowledgments\label{Acknowledgments}}

Portions of analytical calculations were conducted using Wolfram Mathematica. We wish to thank Jeffrey Mills for helpful discussions.

%\pagebreak

\bibliography{mybib}

\begin{thebibliography}{10}
\expandafter\ifx\csname url\endcsname\relax
  \def\url#1{\texttt{#1}}\fi
\expandafter\ifx\csname urlprefix\endcsname\relax\def\urlprefix{URL }\fi
\expandafter\ifx\csname href\endcsname\relax
  \def\href#1#2{#2} \def\path#1{#1}\fi

\bibitem{mcdonald1984some}
J.~B. McDonald, Some generalized functions for the size distribution of income,
  Econometrica 52~(3) (1984) 647--665.

\bibitem{mcdonald1995generalization}
J.~B. McDonald, Y.~J. Xu, A generlazition of the beta distributionwith
  applications, Journal of Econometrics 66 (1996) 133--152.

\bibitem{chotikapanich2008modelling}
D.~Chotikapanjch (Ed.), Modeling Income Distributions and Lorenz Curves,
  Springer, 2008.

\bibitem{chakrabarti2013econophysics}
B.~K. Chakrabarti, A.~Chakraborti, C.~S. R, A.~Chatterjee (Eds.), Econophysics
  of Income and Wealth Distributions,, Cambridge University Press, Cambridge,
  2013.

\bibitem{biewen2018econometrics}
M.~Biewen, E.~Flachaire (Eds.), Econometrics and Income Inequality, 2018.

\bibitem{gomez2018family}
E.~G{\'o}mez-D{\'e}niz, J.~M. Sarabia, A family of generalized beta
  distributions: Properties and applications, Annals of Data Science 5~(3)
  (2018) 401--420.

\bibitem{chotikapanich2018using}
D.~Chotikapanich, W.~E. Griffiths, G.~Hajargasht, W.~Karunarathne, P.~D.~S.
  Rao, Using the gb2 income distribution, Econometrics 6~(2) (2018) 21.

\bibitem{bouchaud2000wealth}
J.-P. Bouchaud, M.~M{\'e}zard, Wealth condensation in a simple model of
  economy, Physica A: Statistical Mechanics and its Applications 282~(3) (2000)
  536--545.

\bibitem{ma2013distribution}
T.~Ma, J.~G. Holden, R.~Serota, Distribution of wealth in a network model of
  the economy, Physica A: Statistical Mechanics and its Applications 392~(10)
  (2013) 2434--2441.

\bibitem{dashti2020stochastic}
M.~Dashti~Moghaddam, J.~Mills, R.~A. Serota, From a stochastic model of
  economic exchange to measures of inequality, Physica A 559 (2020) 125047.

\bibitem{sepplinen2012scaling}
T.~Sepp{\"a}l{\"a}inen, Scaling for a one-dimensional directed polymer with
  boundary conditions, The Annals of Probability 40~(1) (2012) 19--73.

\bibitem{thiery2014loggamma}
T.~Thiery, P.~Le~Doussal, Log-gamma directed polymer with fixed endpoints via
  the replica bethe ansatz, Journal of Statistical Mechanics: Theory and
  Experiment (2014) P10018.

\bibitem{grange2017loggamma}
P.~Grange, Log-gamma directed polymer with one free end via coordinate bethe
  ansatz, Journal of Statistical Mechanics: Theory and Experiment (2017)
  073102.

\bibitem{praetz1972distribution}
P.~D. Praetz, The distribution of share price changes, Journal of Business
  (1972) 49--55.

\bibitem{cox1985theory}
J.~Cox, J.~Ingersoll, S.~Ross, A theory of the term structure of interest
  rates, Econometrica 3~(385--408) (1985).

\bibitem{nelson1990arch}
D.~Nelson, Arch models as diffusion approximations\*, Journal of Econometrics
  45 (1990) 7.

\bibitem{heston1993closed}
S.~L. Heston, A closed-form solution for options with stochastic volatility
  with applications to bond and currency options, The Review of Financial
  Studies 6~(2) (1993) 327--343.

\bibitem{dragulescu2002probability}
A.~A. Dragulescu, V.~M. Yakovenko, Probability distribution of returns in the
  heston model with stochastic volatility, Quantitative Finance 2 (2002)
  445--455.

\bibitem{fuentes2009universal}
M.~A. Fuentes, A.~Gerig, J.~Vicente, Universal behavior of extreme price
  movements in stock markets, PLoS ONE 4~(12) (2009) 1.

\bibitem{ma2014model}
T.~Ma, R.~Serota, A model for stock returns and volatility, Physica A:
  Statistical Mechanics and its Applications 398 (2014) 89--115.

\bibitem{dashti2020modeling}
M.~Dashti~Moghaddam, J.~Liu, J.~G. Holden, R.~Serota, Modeling response time
  distributions with generalized beta prime, Discontinuity, Nonlinearity, and
  Complexity 9~(3) (2020) 477--488.

\bibitem{dashti2021combined}
M.~Dashti~Moghaddam, R.~Serota, Combined mutiplicative-heston model for
  stochastic volatility, Physica A: Statistical Mechanics and its Applications
  561 (2021) 125263.

\bibitem{sornette2009}
D.~Sornette, Dragon-kings, black swans and the prediction of crises,
  International Journal of Terraspace Science and Engineering 2~(1) (2009)
  1--18.

\bibitem{sornette2012dragon}
D.~Sornette, G.~Ouillon, Dragon-kings: Mechanisms, statistical methods and
  empirical evidence, The European Physical Journal Special Topics 205 (2012)
  1--26.

\bibitem{dashti2021realized}
M.~Dashti~Moghaddam, J.~Liu, R.~Serota, Implied and realized volatility: A
  study of distributions and distribution of difference, International Journal
  of Finance and Economics 26 (2021) 2581--2594.

\bibitem{liu2022dragon}
J.~Liu, R.~A. Serota, Are there dragon kings in the stoock market?, in
  preparation (2022).

\bibitem{pisarenko2012robust}
V.~F. Pisarenko, D.~Sornette, Robust statistical tests of dragon-kings beyond
  power law distribution, The European Physical Journal Special Topics 205
  (2012) 95--115.

\bibitem{hertzler2003classical}
G.~Hertzler, "classical" probability distributions for stochastic dynamic
  models, in: 47th Annual Conference of the Australian Agricultural and
  Resource Economics Society, 2003.

\bibitem{eugene2002betanormal}
N.~Eugene, C.~Lee, F.~Famoye, Beta-normal distribution and its applications,
  Communications in Statistics - Theory and Methods 31~(4) (2002) 497--512.

\bibitem{jones2004families}
M.~C. Jones, Families of distributions arising from distributions of order
  statistics, Test Vol. , No. 1, pp. 1 43 13~(1) (2004) 1--43.

\bibitem{cordeiro2009new}
G.~M. Cordeiro, M.~de~Castro, A new family of generalized distributions,
  Journal of Statistical Computation \& Simulation 81~(7) (2009) 883 -- 898.

\bibitem{alexander2012generalized}
C.~Alexander, G.~M. Cordeiro, O.~E.~M. M, J.~M. Sarabia, Generalized
  beta-generated distributions, Computational Statistics and Data Analysis 56
  (2012) 1880--1897.

\bibitem{alzaatrech2013newmethod}
A.~Alzaatrech, C.~Lee, F.~Famoye, A new method for generating families of
  continuous distributions, METRON 71 (2013) 63--79.

\bibitem{lemonte2013extended}
A.~J. Lemonte, G.~M. Cordeiro, An extended lomax distribution, Statistics
  47~(4) (2013) 800--816.

\bibitem{nist2022digital}
Nist digital library of mathematical functions, https://dlmf.nist.gov.

\bibitem{srinivasa2010distance}
S.~Srinivasa, M.~Haenggi, Distance distributions in finite uniformly random
  networks: Theory and applications, IEEE Transactions on Vehicular Technology
  59~(2) (2010) 940--949.

\bibitem{sepanski2007family}
J.~H. Sepanski, L.~Kong, A family of generalized beta distributions for income,
  arXiv:0710.4614 (2007).

\bibitem{risken1996fokker}
H.~Risken, The Fokker-Planck Equation, Springer, 1996.

\bibitem{jacobs2010stochastic}
K.~Jacobs, Stochastic Processes for Physicists, Cambridge University Press,
  2010.

\bibitem{janczura2012black}
J.~Janczura, R.~Weron, Black swans or dragon-kings? a simple test for
  deviations from the power law⋆, European Physical Journal Special Topics
  205 (2012) 79--93.

\bibitem{knuth1998art}
D.~E. Knuth, The Art of Computer Programming, 3rd Edition, Vol.~2, Addison
  Wesley, 1998.

\bibitem{massey1985kolmogorov}
F.~J. Massey, The kolmogorov-smirnov test for goodness of fit, Journal of
  American Statistical Association 80~(392) (1985) 954--958.

\end{thebibliography}

\end{document}